\documentclass[a4paper]{jpconf}
\usepackage{graphicx}
\usepackage{amssymb}

\begin{document}
\title{Reflecting on \v{C}erenkov Reflections }
\author{D. Fargion$^{1,2}$, M. Gaug$^{3}$, P. Oliva$^{1}$\\
\small{$^{1}$ \it{Physics Dep., $^{2}$INFN1, University ``Sapienza'', Ple. A.Moro, 2 00185 Rome, Italy.}\\
$^{3}$ \it{Instituto de Astrof\`{\i}sica de Canarias, Via L\`{a}ctea, 38205 La Laguna, Tenerife, Spain.}}}
\ead{daniele.fargion@roma1.infn.it}
\begin{abstract}
MAGIC, as well as HESS and VERITAS, is a \v{C}erenkov Telescope
unveiling $\gamma$-ray sources above
60 GeV at vertical within noisy (hadronic)
airshowering sky. These telescopes while facing the horizons may
reveal rarest blazing UHECR as well as far fluorescence tails of
downward PeV-EeV hadronic airshowers. Few of these inclined
airshowers blazing on axis are spread by the geomagnetic field into twin
spots. These twin flashes and their morphology may tag the UHECR origination site. There is a rich window of such reflecting \v{C}erenkov lights visible by Telescopes on top of Mountains as MAGIC  (and partially VERITAS): the reflections from the nearby ground (possibly enhanced by rain or snow, ice white cover), from the Sea and from the cloudy sky; in particular, these
cloudy sheets may lay above or below the observer. MAGIC looking
downward to the clouds or the snow, may well reveal blazing Moliere
disks diffusing \v{C}erenkov spots (few events per night).
Because of geomagnetic forces and splitting of the inclined
air-shower, one should reveal for the first time (at tens PeV or above) \v{C}erenkov airshowers whose flashes are skimming the MAGIC nearby Sea and
opened into twin spots. Their morphology may tag the UHECR
origination, its consequent cross-section and  composition. Magic telescopes looking upward into cloudy sky may observe very rare up-going UHE Tau, originated by UHE PeVs neutrinos skimming earth, air-showering into sky, reflecting into clouds. In particular Glashow resonant antineutrinos electron hitting into Earth electrons may lead to gauged boson $W^-$ , whose decay (inside the Earth) may
produce a $\tau+\bar{\nu}_{\tau}$  \cite{Fargion1999}, which later escape and decay in air is producing \v{C}erenkov lights; these flashes may blaze into the clouds above MAGIC as upward dot spots. The Magic energy threshold for such
UHE Neutrinos showers rises to PeV values.  EeV UHE tau neutrinos
by guaranteed  GZK UHECR secondaries \cite{Greisen:1966jv, za66}, via the muon-tau flavor mixing,  may skim the Earth, produce UHE tau particles whose escape in air and decay in flight may blaze Magic or reflect \v{C}erenkov light at opposite far  cloudy sky edges. Any collaboration (MAGIC,HESS,VERITAS) will be  willing to dedicate cloudy time to UHECR physics since nothing else can be done in that time. These  reflections may open a totally new UHECR spectroscopy while unveiling a rare, but loud  Tau Neutrino astronomy\cite{Fargion2007, Fargion1}.
\end{abstract}
\begin{figure}[t]
\includegraphics[width=8.5cm, height=2.8cm]{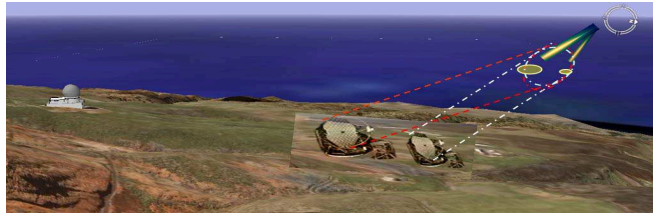}
\centering \caption{Inclined air-showers, charged electron pairs and their \v{C}erenkov lights, split by geomagnetic fields,  are diffused by the
sea and detected as a polarized twin spot on MAGIC telescopes. The scrambling of the waves may diffuse and fragment the ellipse shining areas.} \label{fig1}
\end{figure}
\vspace{-5pt}
\begin{figure}[t]
\includegraphics[width=8.5cm, height=2.8cm]{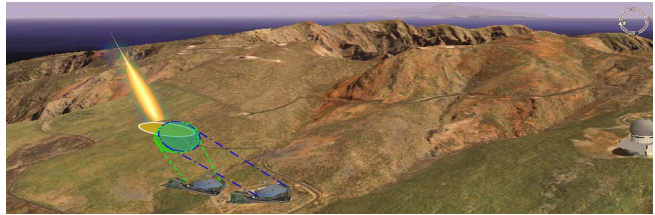}
\centering \caption{A moderate inclined airshower hit the mountain and its \v{C}erenkov light is reflected from the ground to MAGIC telescope which detect a disk (if on axis) or a thin ellipse of diffused light.} \label{fig2}
\end{figure}
\vspace{-14pt}
\begin{figure}[t]
\includegraphics[width=8.5cm, height=2.8cm]{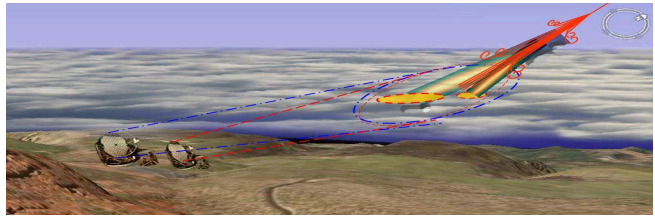}
\centering \caption{Very inclined UHECR air-shower, its \v{C}erenkov splitting lights  is reflected onto clouds below  MAGIC telescope. The twin lobes might
be revealed at once, in rare events. Their separation may inform on
the primary direction, origination, composition.} \label{fig3}
\end{figure}
\begin{figure}[t]
\includegraphics[width=8.5cm, height=2.8cm]{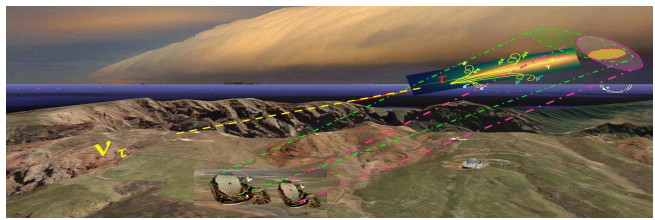}
\centering \caption{\v{C}erenkov lights from up-ward horizontal air-shower
(Hortau-Uptau) shining both fluorescence and \v{C}erenkov reflections onto air and clouds of Magic sky (\cite{Fargion1}, Fig 1). Split shower cones may hardly rise because EeV taus (contrary to hadronic UHECR at horizons) shine mostly at low altitudes, where dense atmosphere suppress geomagnetic separation.}
\label{fig4}
\end{figure}
\section{Blazing and Splitting \v{C}erenkov lights at the edges.}
The common downward vertical airshowers, mostly of hadronic nature, are
growing as a tree whose wide roots are spread in a conical shape, because at low altitudes vertical shower are randomized into a disk. The resultant secondaries shine \v{C}erenkov light on a disk's size which is about twice a Moliere radius (a hundred meter size). At high altitudes, inclined shower (zenith angle $\theta\geqslant70^{\circ}$) at lowest air density are split by geomagnetic Lorentz force into two twin charged beams. At their center, energetic  muon pair bundles define  the shower axis. These  fan-like shape are made by twin  pair lobes.  These lobes are shining \v{C}erenkov lights in different beams and regions; such an exemplar UHECR inclined event has been foreseen in last years,\cite{Fargion3, Fargion4,Fargion5}, recorded by the AUGER detector \cite{Auger07} and discussed recently \cite{Fargion1}: both a \v{C}erenkov flash (seen at Coihueco) and a
fluorescence flare (spot by Los Leones) were detected while a muon
pair bundle hit a dozen of tanks in the heart of the AUGER array (see Fig.\ref{fig2} and Fig.\ref{fig3} \cite{Fargion1}).
Let us remark that only one side (of the two) of an off-axis blazing \v{C}erenkov flash (the beam made by bent negative electrons) have been revealed by Coihueco:
because of the wide fan-like shower size, \v{C}erenkov lights are mostly seen in one main lobe. Only rarely \v{C}erenkov jets are observed in axis being split far away into their twin lobes; the higher energy muon bundles survive, being harder than electron ones, reaching the ground almost undeflected along the main shower axis. Such a downward twin spot discover would offer a new UHECR spectroscopy at the horizon (inclined showers) \cite{Fargion2}.
\v{C}erenkov earliest twin lights define, indeed, the UHECR original direction, its most probable cross-section and its consequent primary composition. \v{C}erenkov light's intensity and color (spectra), zenith angle, traversed column depth, angular separation, time structure and muons component delay and multiplicity, defines primary energy and composition. Moreover, such a \emph{pedagogical}  in-axis event, where the lobes split, even if rare could be enhanced by time correlated muon light rings (by secondary muon \v{C}erenkov imprint \cite{Fargion2, Fargion3, Fargion4, Fargion5}). The MAGIC-II telescope array \cite{ANGELIS}, as well as VERITAS and HESS, should discover such inclined events
if looking horizontally \cite{zas05}. The hope is to prepare at the same time the calibration for rarest but exciting upgoing PeV-EeV neutrino induced airshowers \cite{Fargion1999, Fargion2004}. In the present article we reflect on their Cherenkov reflections summarized by the last four figures.
Let us remind that at highest energies (EeV), Fluorescence and \v{C}erenkov hybrid events at AUGER (estimated about a dozen per year \cite{Fargion1}) may offer an empirical calibration (see Fig.\ref{fig2} and Fig.\ref{fig3}, \cite{Fargion1}). Hybrid events by only \v{C}erenkov twin spots may be extended to lower energies, as tens PeV, at more frequent rate. In analogy, MAGIC may study twin split events: their double spot signature is larger at larger zenith angles though this
separation is observed at more and more far edges. Therefore, the
angular size between the two lobes first  vanishes at vertical than
increases at larger zenith angle to a maximum angle. The \v{C}erenkov lobes'
angular size appears to shrink to smaller
aperture at higher and higher zenith angles, because of their larger
distances at the horizons. The twin signature should be resolved better
by MAGIC telescope than by AUGER because better resolution.
AUGER, on the contrary, may enhance and  tag the twin lobes  by adding
tanks around the four fluorescence-\v{C}erenkov telescopes at
Coihueco, Los Leones, Los Morados and Norte: their trigger
correlated muon signal may, in fact, trigger the reading of
\v{C}erenkov flashes at the edges, revealing  their hidden twin split
signature.
\section{Reflecting on the Sea, ground or clouds around MAGIC.}
While MAGIC observes the far sea, it may reveal reflections of
inclined airshowers on the water: in an ideal flat sea, just a twin
airshower image would appear, see Fig.\ref{fig1}, while in a more realistic scrambled water, any ellipse reflected event would be fragmented and diffused at horizon. The sea water reflectivity increases with zenith angle, but column depth opacity may suppress most of the horizontal showers intensity. Therefore, UHECRs could be indirectly detected by such sea-skimming mostly at bounded inclined zenith angles $ 85^o\geq \theta \geq 70^o$. Stereoscopic detection would better calibrate
the distance, the origin and the primary composition. Polarization
of these mirror images may also test the skimming angle.
A complementary reflecting mode takes place on snow, rainy or bright  areas
nearby the telescope. MAGIC leaping the Earth surface could reveal
PeVs reflected events, in particular if they are beamed and skimming
the ground toward the telescope's direction, see Fig.\ref{fig2}. Moreover, it's  possible to observe at once both the deep downward airshower \v{C}erenkov
reflections and part of the airshower lateral fluorescence tail (see Fig.\ref{fig1} \cite{Fargion1}). The possibility to use a \v{C}erenkov telescope in \emph{reflection mode} may verify known UHECR rates and morphology.
One can image to use the near ground reflections to enhance
the brightness because of the near diffusor distance.
In the skimming case, the spot seen in perspective will appear as an ellipse the more eccentric the more inclined the shower is, but once re-diffused from
ground can reappear on the telescope as a big disk thanks to the
in-axis geometry. Clouds are seldom into sky. High altitude detectors (MAGIC, VERITAS) are often above these clouds: they can work as a blank white screen which could better reveal downward UHECR events, see Fig.\ref{fig3}.
Their splitting may be also revealed. The same clouds blanket above the telescope, when sky is overcast, can be used to diffuse the \v{C}erenkov lights from rare tau upgoing airshower, whose blazing signal may lead to short light spots on such natural huge screens, see Fig.\ref{fig4}  \cite{Fargion1999, Fargion2004, Gaug07}. The areas observed by MAGIC are smaller than the ones of AUGER because of MAGIC smaller solid angle, though longer distances available. However, because of the larger area and smaller energy threshold, MAGIC may reveal PeVs neutrino induced  signals on nearby clouds. One should note that the twin MAGIC view toward (\cite{Fargion2,Becker07}) \emph{and opposite} GRBs-SGRs-BL Lac active $\gamma$ sources direction, while at horizon, can be performed: while one telescope can follow the source beyond the edge, the second telescope could point, in presence of a cloudy sky, the same direction but \emph{opposite to} the source
position. This location is not just a point region but a flat arc,
projected by the horizon edges into cloudy sky, where any UHE
skimming $\tau$ may rise and flash on such a screen. In this working
configuration, blazing \v{C}erenkov flashes (by $\bar{\nu}_e+e\rightarrow W^-$, by $\nu_\tau+X\rightarrow\tau$ or by $\bar{\nu}_\tau+ X\rightarrow\tau$ airshowers), or EeV $\tau$ air-showering may hit one telescope directly while a reflected \v{C}erenkov spots by upgoing $\tau$ showers would be diffused onto the cloud-screen. This can better
occur after GLAST era, when daily GRB events may trigger MAGIC eyes.
Nearly one GRB a month may rise at horizons.
\section{Conclusions: Reflecting Magic lights in dark edges.}
Airshowers \v{C}erenkov lights may be reflected while skimming the
Earth, the Sea, the icy or rainy grounds.The MAGIC Telescopes (as
well as the VERITAS ones) being on top of a mountain, may soon reveal these
reflections whose role is also to calibrate the downward UHECR rate.
In principle, MAGIC telescope observing large zenith angles, may also
search for upgoing tau airshower probably related to GRB-SGRs
or BL Lac brightening. Also Glashow resonant neutrinos may rise
upgoing and blazing airshowers at horizons. Because of the atmosphere dimming of
far hadronic airshowers at Earth atmosphere edges, the nearer \cite{Fargion1999}, younger \cite{Bertou2002}neutrino induced ones are emerging better,  over the far hadronic noise. In particular the Earth skimming Hortaus (horizontal $\tau$s) \cite{Fargion2004} while decaying and showering may, at the same time, blaze at the front and-or  flash on the
opposite side to the (eventual) clouds or by fluorescence in air. The location where these reflection might hit is not a point anti-correlated to the source but a wide  arc parallel to the horizon line. All the hadronic inclined
air-showers may and must show the geomagnetic splitting whose signature can be
a test of UHECR  rate and composition from the knee up to GZK energy edges. Just where AUGER most recent and surprising results on UHECR  call for a confirmation \cite{Yamamoto2007}. In larger AUGER sky, similar higher energy UHE tau events (possibly originated by EeV GZK \cite{Greisen:1966jv, za66} tau neutrinos),  events whose spectra and local anisotropy has been recently revealed, must shine in next few years\cite{Fargion2007}.

\end{document}